\documentclass[12pt]{article}
\usepackage{amsmath}
\usepackage{amsfonts}
\usepackage{amssymb}
\usepackage{graphicx}
\usepackage{color}
\usepackage[table,xcdraw]{xcolor}
\usepackage{orcidlink}
\usepackage{hyperref}
\hypersetup{colorlinks=true, linkcolor=blue, citecolor=blue, urlcolor=blue}
\usepackage{xcolor}
\usepackage{xstring}
\newcommand{\hrefbibentry}[2]{%
  \IfStrEq{#1}{}{#2}{%
    \IfBeginWith{#1}{http}%
      {\href{#1}{\textcolor{blue}{#2}}}%
      {\href{https://doi.org/#1}{\textcolor{blue}{#2}}}}}
\usepackage{float}
\usepackage{multirow}
\usepackage{array}
\usepackage{longtable}
\usepackage{booktabs}
\usepackage{colortbl}
\usepackage{tabularx}
\usepackage{xltabular}
\newcolumntype{C}{>{\centering\arraybackslash}X}
\newcolumntype{L}{>{\raggedright\arraybackslash}X}
\newcolumntype{R}{>{\raggedleft\arraybackslash}X}
\usepackage{geometry}
\RequirePackage[numbers,sort&compress]{natbib}
\numberwithin{equation}{section}
\usepackage{doi}

\newcommand{\rh}{r_{h}}
\newcommand{\VYM}{V_{\rm YM}}
\newcommand{\Vh}{\widehat{V}_{\rm YM}}
\allowdisplaybreaks
\begin{document}
\baselineskip=20pt

\begin{center}
\LARGE{Thermodynamics, topology and non-Abelian instability of the
five-dimensional Einstein--Yang--Mills black hole with exponential entropy
correction}
\end{center}
\vspace{0.3cm}

\begin{center}
{\bf Aram Bahroz Brzo\orcidlink{0000-0002-1257-9377}}\footnote{\bf aram.brzo@univsul.edu.iq}$^{a}$,\ \
{\bf Peshwaz Abdulkareem Abdoul\orcidlink{0000-0002-2144-8336}}\footnote{\bf peshwaz.abdoul@chu.edu.iq}$^{b}$,\ \
{\bf Behnam Pourhassan\orcidlink{0000-0003-1338-7083}}\footnote{\bf b.pourhassan@du.ac.ir}$^{c,d}$\ \ and\ \
{\bf \.{I}zzet Sakall{\i}\orcidlink{0000-0001-7827-9476}}\footnote{\bf izzet.sakalli@emu.edu.tr}$^{e}$\\[6pt]
{\it $^{a}$Physics Department, College of Education, University of Sulaimani, Sulaimani 46001, Kurdistan Region, Iraq}\\
{\it $^{b}$Physics Department, College of Science, Charmo University, Chamchamal, Sulaimani, Kurdistan Region, Iraq}\\
{\it $^{c}$School of Physics, Damghan University, P.~O.~Box 3671641167, Damghan, Iran}\\
{\it $^{d}$Center for Theoretical Physics, Khazar University, 41 Mehseti Street, Baku, AZ1096, Azerbaijan}\\
{\it $^{e}$Physics Department, Eastern Mediterranean University, Famagusta 99628, North Cyprus via Mersin 10, T\"{u}rkiye}
\end{center}
\vspace{0.3cm}

\abstract{\noindent
The five-dimensional Einstein--Yang--Mills  black hole obtained from the
higher-dimensional Wu--Yang ansatz carries a logarithmic non-Abelian
contribution to the metric function, which makes it an analytically tractable
setting for studying how gauge charge reshapes horizon thermodynamics beyond
four dimensions. We derive closed-form expressions for the mass--radius
relation, the Hawking temperature, the entropy, the Yang--Mills potential and
the first law, and we record the curvature invariants, which show that the
logarithm sharpens the central singularity rather than smoothing it. A
non-perturbative exponential entropy correction is then added, and the
corrected heat capacity, Helmholtz free energy and Gibbs-like potential are
obtained analytically. The geometry is extremal at $\rh=|Q|$ and the heat
capacity diverges at $\rh=\sqrt{3}\,|Q|$, a Davies-type point. The correction
moves that point inward,
$r_{\rm tr}/|Q|\simeq\sqrt{3}-9\lambda\alpha\eta\,e^{-3\sqrt{3}\lambda}$.
Using the generalized off-shell free energy we find a conserved topological
charge $W=0$ carried by a stable--unstable defect pair whose inner member alone
responds to the correction. We then solve the spherically symmetric magnetic
Yang--Mills perturbation sector, which probes the gauge field itself rather
than a test scalar. Two independent numerical schemes agree to better than
$1.2\times10^{-7}$ and give a purely imaginary unstable mode whose growth rate
falls monotonically from $\Gamma_{0}\rh=0.414753$ at $Q/\rh\to0$ to $0.317520$
at $Q/\rh=0.99$, passing through the Davies scale without any feature. Because
the effective potential decays as $-9/(4r_{*}^{2})$, with a coefficient above
the critical value $1/4$, the sector supports an infinite geometric tower of
unstable modes accumulating at $\Gamma\to0$ with the charge-independent ratio
$\Gamma_{n+1}/\Gamma_{n}\to e^{-\pi/\sqrt{2}}\simeq0.108453$, which we confirm
numerically.}

\vspace{0.4cm}
\noindent\textbf{Keywords:} five-dimensional black holes; Einstein--Yang--Mills
theory; Wu--Yang ansatz; exponential entropy correction; thermodynamic
topology; quasinormal modes; non-Abelian instability

\vspace{0.4cm}
\section{Introduction}\label{isec1}

Einstein--Yang--Mills (EYM) theory is the most direct nonlinear extension of
Einstein gravity coupled to a gauge sector. Yang--Mills fields are non-Abelian,
so their generators fail to commute and the gauge sector self-interacts even
without external sources. That single feature separates EYM gravity from
Einstein--Maxwell theory and controls much of what follows for black-hole
physics, gravitational collapse and horizon thermodynamics
\cite{Yasskin1975,BartnikMcKinnon1988,Bizon1990,Smoller1993}.

The early history is instructive. Yasskin showed that Einstein--Maxwell
solutions can be embedded in EYM theory whenever the gauge group admits an
invariant metric \cite{Yasskin1975}, which built a bridge between the Abelian
and non-Abelian sectors. Bartnik and McKinnon then found globally regular
particle-like solutions with no Maxwell counterpart \cite{BartnikMcKinnon1988},
and Bizo\'n introduced colored black holes, whose branch structure and
stability differ sharply from Reissner--Nordstr\"om \cite{Bizon1990}. Existence
proofs followed from Smoller, Wasserman and Yau \cite{Smoller1993}. Stability
turned out to be the delicate part: the colored solutions were shown to be
unstable \cite{StraumannZhou1990,Bizon1991}, and the argument was later
extended to arbitrary gauge groups \cite{BrodbeckStraumann1996}. Instability is
thus the rule rather than the exception in this class of solutions, a point we
return to in Sec.~\ref{isec6}.

Higher-dimensional gravity developed in parallel. String theory, supergravity
and braneworld constructions all pushed in that direction. Tangherlini
generalized the
Schwarzschild geometry to arbitrary dimension \cite{Tangherlini1963}, and
higher-dimensional horizons now appear throughout gauge/gravity duality,
black-brane physics and extended thermodynamics \cite{Wald2001}. Higher-dimensional
Hawking emission is itself sensitive to the nonlinearity of the matter sector
\cite{MazharimousaviSakalliHalilsoy2009}, and the same sensitivity carries over
to the bumblebee and other Lorentz-violating extensions in dimension greater
than four \cite{UniyalKanziSakalli2023}. It is natural to ask what a non-Abelian
gauge field does to a horizon in five dimensions.

Mazharimousavi and Halilsoy answered part of that question by constructing
exact higher-dimensional EYM black holes from a higher-dimensional Wu--Yang
ansatz \cite{MazharimousaviHalilsoy2008}. In five dimensions the metric
function closes in a short form containing a logarithm of the areal radius.
The logarithm is the whole point. It is not the Maxwell scaling one would
expect in five dimensions, and it gives the horizon and its thermodynamics a
structure of their own \cite{MazharimousaviHalilsoy2007}. The solution has
continued to attract attention in studies of quantum-corrected thermodynamics
\cite{Rather2025}, and related Einstein--Power--Yang--Mills geometries have
been examined through tunneling and orbital observables
\cite{SucuSakalli2025EPYM}.

Black-hole thermodynamics supplies the language. The four laws were set out by
Bardeen, Carter and Hawking \cite{BardeenCarterHawking1973}, Bekenstein tied
entropy to horizon area \cite{Bekenstein1973}, and Hawking gave the temperature
its physical meaning \cite{Hawking1975,Hawking1976}. Beyond Einstein gravity
the entropy is read off from Wald's Noether charge \cite{Wald1993,IyerWald1994}.
The same machinery has been applied to horizons dressed by matter sectors of
various kinds, from Lorentz-violating backgrounds \cite{SucuSakalli2025KR} to
regular geometries surrounded by dark matter and string clouds
\cite{AlBadawiAhmedSakalli2026}.
Quantum corrections to that entropy come in two families. Logarithmic terms
arise at one loop and are well understood \cite{aram}. Non-perturbative
exponential terms are the more recent development
\cite{Dabholkar1995,ChatterjeeGhosh2020,PourhassanFaizal2021,PourhassanEtAl2024}:
they are negligible for large horizons and can dominate for small ones, which
makes them attractive whenever the near-extremal sector is the physically
interesting one \cite{aram1,aram2,aram3,aram4,aram5,aram6}. Exponentially
modified entropies have also been extracted from tunneling calculations with
generalized uncertainty corrections
\cite{SucuSakalliSucu2026PLB,SucuSakalliSucu2026AdP}. The five-dimensional EYM
black hole has an extremal radius available in closed form, so it is a clean
place to ask what such corrections do to the heat capacity, the free energy and
the phase portrait.

Thermodynamic topology is the second ingredient. The phase structure is encoded
in the zeros of a vector field built from a generalized off-shell free energy,
and each zero carries a winding number contributing to a total topological
charge \cite{DuanFuJia2000,WeiLiuMann2022,WeiLiuWang2022,Alipour2023,AliEtAl2024}.
The classification is global, which is what makes it more informative than the
local sign of the heat capacity, and it has been applied to non-extensive and
quantum-corrected settings alike
\cite{GashtiPourhassanSakalli2025,GashtiSakalliPourhassan2025,aram3,aram4,aram5,aram6}.
The five-dimensional EYM geometry is simple enough to be handled by hand and
still rich enough to produce a nontrivial defect structure.

Quasinormal modes (QNMs) supply the dynamical probe, with complex frequencies
fixed by ingoing behavior at the horizon and outgoing behavior far away
\cite{Nollert1999,KokkotasSchmidt1999,BertiCardosoStarinets2009,KonoplyaZhidenko2011}.
Higher-dimensional applications are standard, including gauge-invariant
treatments of gravitational perturbations
\cite{KodamaIshibashi2003,KodamaIshibashi2011} and higher-order WKB spectra of
$D$-dimensional Schwarzschild geometries \cite{Konoplya2003}. Regge--Wheeler
master equations have likewise been adapted to quantum-corrected backgrounds
carrying topological charge \cite{AhmedAlBadawiSakalli2026}. For the specific
five-dimensional EYM black hole studied here, Guo and Miao already computed the
scalar spectrum and its dependence on the Yang--Mills charge \cite{GuoMiao2020},
and the eikonal relation to unstable null geodesics was worked out shortly
afterwards \cite{GuoMiao2020b}. We therefore do not repeat the scalar
calculation.

What is missing is a dynamical test of the gauge field itself. A probe scalar
on a fixed background says nothing about whether the Yang--Mills configuration
holds together, and the four-dimensional experience
\cite{StraumannZhou1990,Bizon1991,BrodbeckStraumann1996} suggests it may not.
We therefore study the spherically symmetric magnetic Yang--Mills perturbation
sector, following the five-dimensional framework of Okuyama and Maeda
\cite{OkuyamaMaeda2003}. We derive the master equation and effective potential
for the logarithmic geometry, solve the resulting eigenvalue problem with two
independent numerical methods, and compare the instability rate against the
thermodynamic scales $\rh=|Q|$ and $\rh=\sqrt{3}\,|Q|$. Along the way we find
that the asymptotic form of the effective potential forces an infinite tower of
unstable modes, a structural feature that does not seem to have been noted for
this solution.

The aim of this work is therefore not to revisit a known geometry but to place
four things side by side for the same background: exact classical
thermodynamics, a non-perturbative entropy correction, a global topological
classification, and the intrinsic stability of the non-Abelian sector.

Section~\ref{isec2} presents the solution and its curvature invariants.
Section~\ref{isec3} covers the classical thermodynamics. The exponential
entropy correction and the shifted Davies transition occupy
Sec.~\ref{isec4}, and Sec.~\ref{isec5} builds the topological description.
Section~\ref{isec6} solves the magnetic Yang--Mills sector, and
Sec.~\ref{isec7} collects the discussion and conclusions. We work in units
where $c=\hbar=k_{B}=1$ and keep the five-dimensional Newton constant $G_{5}$
explicit.

\section{Geometry of the five-dimensional Einstein--Yang--Mills black hole}
\label{isec2}

The five-dimensional EYM action reads
\cite{Yasskin1975,MazharimousaviHalilsoy2008,Rather2025}
\begin{equation}
I=\frac{1}{16\pi G_{5}}\int d^{5}x\,\sqrt{-g}\,
\left(R-\mathcal{F}\right),
\label{eq:action}
\end{equation}
where $R$ is the Ricci scalar and $\mathcal{F}$ is the Yang--Mills invariant
built from the non-Abelian field strength
\begin{equation}
F^{(a)}_{\mu\nu}=\partial_{\mu}A^{(a)}_{\nu}-\partial_{\nu}A^{(a)}_{\mu}
+\frac{1}{\sigma}\,C^{(a)}_{\phantom{(a)}(b)(c)}A^{(b)}_{\mu}A^{(c)}_{\nu}.
\label{eq:fieldstrength}
\end{equation}
Here $A^{(a)}_{\mu}$ is the gauge potential, $\sigma$ the coupling constant and
$C^{(a)}_{\phantom{(a)}(b)(c)}$ the structure constants of the gauge group
\cite{Yasskin1975,MazharimousaviHalilsoy2008}. In the higher-dimensional
Wu--Yang construction the gauge field is purely magnetic, which reduces the
field equations to a spherically symmetric system that can be integrated
exactly \cite{MazharimousaviHalilsoy2008,MazharimousaviHalilsoy2007}.

For the static line element
\begin{equation}
ds^{2}=-f(r)\,dt^{2}+\frac{dr^{2}}{f(r)}+r^{2}\,d\Omega_{3}^{2},
\qquad \Omega_{3}=2\pi^{2},
\label{eq:metric}
\end{equation}
with $d\Omega_{3}^{2}$ the metric on the unit three-sphere, the exact solution
is \cite{MazharimousaviHalilsoy2008,MazharimousaviHalilsoy2007,Rather2025}
\begin{equation}
f(r)=1-\frac{m}{r^{2}}-\frac{2Q^{2}\ln r}{r^{2}},
\label{eq:f}
\end{equation}
where $m$ is the integration constant and $Q$ the Yang--Mills magnetic charge.
Setting $Q=0$ returns the Schwarzschild--Tangherlini geometry. The logarithm is
what distinguishes the non-Abelian solution from its Maxwell counterpart in
five dimensions \cite{MazharimousaviHalilsoy2008}.

The event horizon follows from $f(\rh)=0$, which fixes
\begin{equation}
m=\rh^{2}-2Q^{2}\ln \rh,
\label{eq:m}
\end{equation}
and eliminating $m$ in favor of $\rh$ gives the form used throughout,
\begin{equation}
f(r)=1-\frac{1}{r^{2}}\left[\rh^{2}+2Q^{2}\ln\!\left(\frac{r}{\rh}\right)\right].
\label{eq:fh}
\end{equation}
With the Tangherlini normalization $M=3\pi m/(8G_{5})$
\cite{Tangherlini1963,Wald2001}, the mass--radius relation becomes
\begin{equation}
M(\rh,Q)=\frac{3\pi}{8G_{5}}\left(\rh^{2}-2Q^{2}\ln \rh\right).
\label{eq:M}
\end{equation}
Equation~\eqref{eq:M} carries the entire thermodynamic analysis that follows.

Before turning to thermodynamics it is worth recording what the curvature does.
Direct computation from Eq.~\eqref{eq:metric} gives
\begin{equation}
R=\frac{2Q^{2}}{r^{4}},
\label{eq:ricci}
\end{equation}
so the Ricci scalar is sourced entirely by the gauge field and vanishes in the
Tangherlini limit, as it must, since the five-dimensional Yang--Mills stress
tensor is not traceless. For the Kretschmann invariant we find the compact
result
\begin{equation}
K=R_{\mu\nu\rho\sigma}R^{\mu\nu\rho\sigma}
=\frac{72\,h^{2}-168\,Q^{2}h+124\,Q^{4}}{r^{8}},
\qquad h(r)\equiv m+2Q^{2}\ln r,
\label{eq:kretschmann}
\end{equation}
which reduces to the known $72m^{2}/r^{8}$ when $Q=0$. Since
$h\to-\infty$ logarithmically as $r\to0$, the leading behavior is
\begin{equation}
K\simeq\frac{288\,Q^{4}\ln^{2}r}{r^{8}},\qquad r\to0.
\label{eq:Kasym}
\end{equation}
The origin therefore remains a curvature singularity, and the non-Abelian
charge strengthens the divergence by a squared logarithm relative to
Tangherlini rather than smoothing it. No regular core is generated by the
Wu--Yang gauge field, which is worth stating explicitly because the logarithm
in Eq.~\eqref{eq:f} is sometimes read as a softening of the short-distance
behavior.

\section{Classical thermodynamics}\label{isec3}

The Hawking temperature follows from the surface gravity. Since
$g_{tt}g_{rr}=-1$ for the metric \eqref{eq:metric}, the standard relation
$T=f'(\rh)/4\pi$ applies without modification, and Eq.~\eqref{eq:fh} yields
\begin{equation}
T(\rh,Q)=\frac{\rh^{2}-Q^{2}}{2\pi \rh^{3}}.
\label{eq:T}
\end{equation}
The result is short enough to read off directly: the Yang--Mills charge lowers
the temperature, and the horizon becomes degenerate when the numerator
vanishes. Setting $T=0$ gives the extremal radius
\begin{equation}
r_{\star}=|Q|,
\label{eq:rext}
\end{equation}
in closed form, which is unusual among higher-curvature and regular black-hole
models. Substituting Eq.~\eqref{eq:rext} into Eq.~\eqref{eq:M},
\begin{equation}
M_{\star}=\frac{3\pi Q^{2}}{8G_{5}}\left(1-2\ln|Q|\right).
\label{eq:Mext}
\end{equation}

The Bekenstein--Hawking entropy is the horizon area over $4G_{5}$
\cite{Bekenstein1973,Wald2001}. In five dimensions
$A=\Omega_{3}\rh^{3}=2\pi^{2}\rh^{3}$, so
\begin{equation}
S_{0}=\frac{\pi^{2}\rh^{3}}{2G_{5}},
\qquad
\frac{dS_{0}}{d\rh}=\frac{3\pi^{2}\rh^{2}}{2G_{5}}.
\label{eq:S0}
\end{equation}
Differentiating Eq.~\eqref{eq:M} at fixed charge,
\begin{equation}
\frac{dM}{d\rh}=\frac{3\pi}{4G_{5}}\,\frac{\rh^{2}-Q^{2}}{\rh},
\label{eq:dM}
\end{equation}
and combining Eqs.~\eqref{eq:T} and \eqref{eq:S0} reproduces exactly the same
expression, so
\begin{equation}
\frac{dM}{d\rh}=T\,\frac{dS_{0}}{d\rh},
\qquad\Longrightarrow\qquad
dM=T\,dS_{0}+\Phi_{\rm YM}\,dQ.
\label{eq:firstlaw}
\end{equation}
The first law holds identically, which is a useful consistency check on the
normalization of Eq.~\eqref{eq:M}. The conjugate Yang--Mills potential follows
from Eq.~\eqref{eq:M} at fixed $S_{0}$, equivalently at fixed $\rh$,
\begin{equation}
\Phi_{\rm YM}(\rh,Q)=\left(\frac{\partial M}{\partial Q}\right)_{S_{0}}
=-\frac{3\pi Q\ln \rh}{2G_{5}},
\label{eq:Phi}
\end{equation}
and measures how the mass responds to a change in non-Abelian charge.

The heat capacity at fixed charge is obtained from Eqs.~\eqref{eq:T} and
\eqref{eq:dM},
\begin{equation}
C_{Q}=\left(\frac{dM/d\rh}{dT/d\rh}\right)_{Q}
=-\frac{3\pi^{2}\rh^{3}\left(\rh^{2}-Q^{2}\right)}
{2G_{5}\left(\rh^{2}-3Q^{2}\right)},
\qquad
\frac{dT}{d\rh}=\frac{3Q^{2}-\rh^{2}}{2\pi \rh^{4}}.
\label{eq:CQ}
\end{equation}
Two radii organize the local stability. The factor $\rh^{2}-Q^{2}$ makes
$C_{Q}$ vanish at the extremal point $\rh=|Q|$, and the factor
$\rh^{2}-3Q^{2}$ makes it diverge at
\begin{equation}
\rh=\sqrt{3}\,|Q|,
\label{eq:davies}
\end{equation}
where the temperature reaches its maximum. This is a Davies-type transition
point \cite{Davies1989}. Across it the sign of $C_{Q}$ flips: the branch
$|Q|<\rh<\sqrt{3}|Q|$ is locally stable, and the large-horizon branch
$\rh>\sqrt{3}|Q|$ is locally unstable, as expected for an asymptotically flat
geometry. The physical reading of such divergences has been revisited recently
and need not signal a genuine phase transition \cite{Davies2024}, which is one
reason for supplementing the local analysis with the topological description of
Sec.~\ref{isec5}.

The Helmholtz free energy at fixed charge is
\begin{equation}
F=M-TS_{0}
=\frac{\pi}{8G_{5}}\left[\rh^{2}+2Q^{2}\left(1-3\ln \rh\right)\right],
\label{eq:F}
\end{equation}
with $F<0$ marking the thermodynamically favored branch. Because the system
carries a non-Abelian charge, it is also useful to remove the Yang--Mills work
term and define the Gibbs-like potential
\begin{equation}
G_{\rm YM}=M-TS_{0}-\Phi_{\rm YM}Q
=\frac{\pi}{8G_{5}}\left[\rh^{2}+2Q^{2}\left(1+3\ln \rh\right)\right],
\label{eq:G}
\end{equation}
where the sign of the logarithmic term has flipped relative to
Eq.~\eqref{eq:F} because $-\Phi_{\rm YM}Q$ contributes
$+3\pi Q^{2}\ln \rh/(2G_{5})$. Equation~\eqref{eq:G} is the appropriate
potential in the ensemble where $\Phi_{\rm YM}$, rather than $Q$, is held
fixed.

\section{Exponential entropy correction}\label{isec4}

To reach the small-horizon regime we add a non-perturbative correction written
as a function of the classical entropy itself
\cite{ChatterjeeGhosh2020,PourhassanFaizal2021,PourhassanEtAl2024},
\begin{equation}
S_{\eta}(\rh)=S_{0}+\eta\,e^{-\alpha S_{0}}
=\frac{\pi^{2}\rh^{3}}{2G_{5}}
+\eta\exp\!\left(-\frac{\alpha\pi^{2}\rh^{3}}{2G_{5}}\right),
\label{eq:Seta}
\end{equation}
where $\eta$ sets the strength of the correction and $\alpha>0$ the suppression
scale. Differentiating,
\begin{equation}
\frac{dS_{\eta}}{d\rh}=\frac{dS_{0}}{d\rh}
\left[1-\alpha\eta\,e^{-\alpha S_{0}}\right]
\equiv\frac{dS_{0}}{d\rh}\,\mathcal{D}(\rh),
\label{eq:dSeta}
\end{equation}
which makes the structure transparent. For large $\rh$ the exponential is
crushed and $S_{\eta}\to S_{0}$; only near extremality, where the horizon is
small, does the correction matter.

We neglect back-reaction, so the mass function is unchanged and the corrected
temperature follows from Eqs.~\eqref{eq:dM} and \eqref{eq:dSeta},
\begin{equation}
T_{\eta}(\rh)=\frac{dM}{dS_{\eta}}
=\frac{\rh^{2}-Q^{2}}{2\pi \rh^{3}}\,
\frac{1}{1-\alpha\eta\,e^{-\alpha S_{0}}}.
\label{eq:Teta}
\end{equation}
As $\alpha\eta\to0$ this collapses to Eq.~\eqref{eq:T}. Its derivative is
\begin{equation}
\frac{dT_{\eta}}{d\rh}
=\frac{1}{2\pi}\left(\frac{3Q^{2}}{\rh^{4}}-\frac{1}{\rh^{2}}\right)
\frac{1}{\mathcal{D}}
-\frac{3\alpha^{2}\pi\eta}{4G_{5}}\,
\frac{\left(\rh^{2}-Q^{2}\right)e^{-\alpha S_{0}}}{\rh\,\mathcal{D}^{2}},
\label{eq:dTeta}
\end{equation}
and the corrected heat capacity is
\begin{equation}
C_{\eta}=T_{\eta}\left(\frac{dS_{\eta}}{dT_{\eta}}\right)_{Q}
=\frac{3\pi\left(\rh^{2}-Q^{2}\right)}{4G_{5}\,\rh}
\left(\frac{dT_{\eta}}{d\rh}\right)^{-1},
\label{eq:Ceta}
\end{equation}
where the prefactor is just $dM/d\rh$, since the factor $\mathcal{D}$ cancels
between $T_{\eta}$ and $dS_{\eta}/d\rh$. Positive $C_{\eta}$ marks a locally
stable branch, negative $C_{\eta}$ an unstable one, zeros mark extremal or
branch-ending states, and poles are candidate second-order transitions. The
corrected free energies follow immediately,
\begin{equation}
F_{\eta}=M-T_{\eta}S_{\eta},
\qquad
G_{\eta}=M-T_{\eta}S_{\eta}-\Phi_{\rm YM}Q,
\label{eq:Feta}
\end{equation}
with $M$ and $\Phi_{\rm YM}$ untouched by the correction because
back-reaction has been dropped.

\begin{figure}[ht!]
\centering
\includegraphics[width=\textwidth]{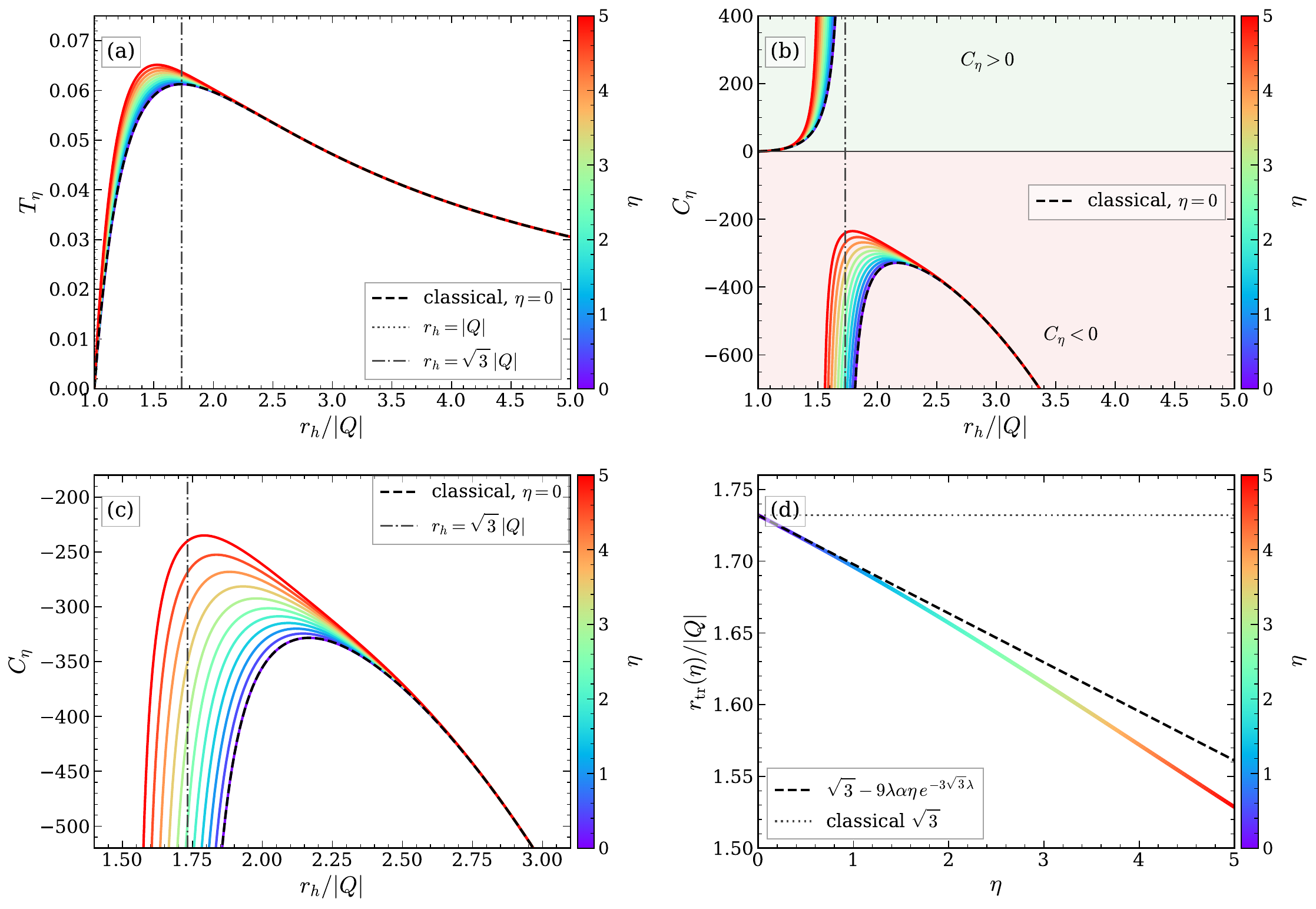}
\caption{Effect of the exponential entropy correction at $G_{5}=1$, $Q=1$ and
$\alpha=0.1$, with the correction strength $\eta$ carried by the colour bar and
the classical limit shown as a dashed black curve in every panel. Panel (a) is
the corrected Hawking temperature $T_{\eta}$ of Eq.~\eqref{eq:Teta} against the
normalized horizon radius, with the extremal radius $\rh=|Q|$ and the classical
Davies scale $\rh=\sqrt{3}|Q|$ marked. Panel (b) is the corrected heat capacity
$C_{\eta}$ of Eq.~\eqref{eq:Ceta}, with the locally stable ($C_{\eta}>0$) and
unstable ($C_{\eta}<0$) regions shaded; the two branches are separated by the
pole at the corrected transition radius. Panel (c) resolves the negative branch
alone, over the reduced range $1.4\le \rh/|Q|\le3.1$ that this branch actually
occupies, and shows how its extremum rises with $\eta$. Panel (d) gives the
transition radius $r_{\rm tr}(\eta)/|Q|$ obtained from
Eq.~\eqref{eq:transcendental}, compared with the linear estimate
\eqref{eq:shift} and with the classical value $\sqrt{3}$. Increasing $\eta$
moves the transition inward and leaves the large-horizon thermodynamics
untouched.}
\label{fig:thermo}
\end{figure}

The Davies-type transition of the corrected theory sits at the extremum of
$T_{\eta}$, equivalently at the pole of $C_{\eta}$. Introducing
\begin{equation}
x=\frac{\rh}{|Q|},
\qquad
\lambda=\frac{\alpha\pi^{2}|Q|^{3}}{2G_{5}},
\label{eq:dimensionless}
\end{equation}
the condition $dT_{\eta}/d\rh=0$ becomes the transcendental equation
\begin{equation}
\frac{3-x_{\rm tr}^{2}}{x_{\rm tr}\left(x_{\rm tr}^{2}-1\right)}
=\frac{3\lambda\alpha\eta\,x_{\rm tr}^{2}\,e^{-\lambda x_{\rm tr}^{3}}}
{1-\alpha\eta\,e^{-\lambda x_{\rm tr}^{3}}}.
\label{eq:transcendental}
\end{equation}
For $\eta\to0$ the right-hand side vanishes and $x_{\rm tr}=\sqrt{3}$ is
recovered. Expanding about that point for weak correction gives
\begin{equation}
\frac{r_{\rm tr}}{|Q|}\simeq\sqrt{3}
-9\lambda\alpha\eta\,e^{-3\sqrt{3}\lambda}+\mathcal{O}(\eta^{2}),
\label{eq:shift}
\end{equation}
so for positive $\alpha$ and $\eta$ the transition moves inward. Numerically,
at $G_{5}=1$, $Q=1$ and $\alpha=0.1$ one has $\lambda\simeq0.49348$, and
Eq.~\eqref{eq:shift} tracks the exact root of
Eq.~\eqref{eq:transcendental} to within $0.1\%$ at $\eta=1$, degrading to about
$2\%$ at $\eta=5$ as the quadratic term takes hold. Table~\ref{tab:davies}
lists the exact roots, the corrected temperature at the transition, and the
relative shift $\Delta r_{D}/r_{D}^{(0)}$. Figure~\ref{fig:thermo} shows the
corrected temperature, both branches of the corrected heat capacity, and the
displacement of the transition radius with $\eta$.

\begin{table}[ht!]
\centering
\caption{Quantum-corrected transition data at $G_{5}=1$, $Q=1$ and
$\alpha=0.1$. The Davies-type radius $r_{D}$ is the root of
$dT_{\eta}/d\rh=0$, equivalently the pole of $C_{\eta}$, and $T_{D}$ is the
corrected temperature evaluated there. The classical limit $\eta=0$ returns
$r_{D}/|Q|=\sqrt{3}$. The relative shift is
$\Delta r_{D}/r_{D}^{(0)}=[r_{D}(\eta)-r_{D}(0)]/r_{D}(0)$, and the last
column is the linear estimate \eqref{eq:shift}.}
\label{tab:davies}
\begin{tabularx}{\textwidth}{CCCCC}
\hline\hline
\textbf{$\eta$} & \textbf{$r_{D}/|Q|$} & \textbf{$T_{D}$} &
\textbf{$\Delta r_{D}/r_{D}^{(0)}$} & \textbf{Eq.~\eqref{eq:shift}}\\
\hline
0 & 1.732051 & 0.0612588 & $0$                       & 1.732051\\
1 & 1.696134 & 0.0617732 & $-2.0737\times10^{-2}$    & 1.697861\\
2 & 1.656968 & 0.0623890 & $-4.3349\times10^{-2}$    & 1.663671\\
3 & 1.615229 & 0.0631337 & $-6.7447\times10^{-2}$    & 1.629482\\
4 & 1.572021 & 0.0640406 & $-9.2393\times10^{-2}$    & 1.595292\\
5 & 1.528617 & 0.0651485 & $-1.1745\times10^{-1}$    & 1.561102\\
\hline\hline
\end{tabularx}
\end{table}

\section{Thermodynamic topology}\label{isec5}

For the global picture we follow the defect construction of
Refs.~\cite{DuanFuJia2000,WeiLiuMann2022,WeiLiuWang2022,Alipour2023,AliEtAl2024}
and define the generalized off-shell free energy
\begin{equation}
\mathcal{F}_{\eta}(\rh,\tau)=M(\rh,Q)-\tau\,S_{\eta}(\rh),
\label{eq:offshell}
\end{equation}
with $\tau$ an auxiliary inverse-temperature parameter. The associated vector
field is
\begin{equation}
\phi=\left(\phi^{\rh},\phi^{\Theta}\right)
=\left(\frac{\partial \mathcal{F}_{\eta}}{\partial \rh},
-\cot\Theta\,\csc\Theta\right),
\label{eq:vectorfield}
\end{equation}
where $\Theta$ compactifies the thermodynamic parameter space. The second
component vanishes only at $\Theta=\pi/2$, so the defects sit on that line and
their radial positions solve $\partial\mathcal{F}_{\eta}/\partial \rh=0$. Using
Eqs.~\eqref{eq:dM} and \eqref{eq:dSeta} and solving for $\tau$ gives the defect
curve
\begin{equation}
\tau(\rh)=\frac{\rh^{2}-Q^{2}}{2\pi \rh^{3}
\left[1-\alpha\eta\,e^{-\alpha S_{0}}\right]}.
\label{eq:defect}
\end{equation}
Comparison with Eq.~\eqref{eq:Teta} shows that $\tau(\rh)=T_{\eta}(\rh)$
identically: the defect curve is the corrected temperature, and the zeros at a
given $\tau$ are the horizon radii whose corrected temperature equals the
auxiliary parameter. This identification is what ties the topological data to
the local stability analysis of Sec.~\ref{isec4} without further input.

The same two factors organize the topology as organized the heat capacity. The
combination $\rh^{2}-Q^{2}$ appears in the temperature, in the mass derivative,
in the corrected heat capacity and in Eq.~\eqref{eq:defect}, so the extremal
point $\rh=|Q|$ is simultaneously the zero-temperature state, the zero of both
heat capacities and a distinguished endpoint of the defect curve. The
combination $\rh^{2}-3Q^{2}$ controls the pole, so $\rh=\sqrt{3}|Q|$ sets the
transition scale. The Yang--Mills charge is thus not merely an extra conserved
quantity; it fixes the scale of both the local and the global phase structure.

Each isolated zero carries the winding number
\begin{equation}
w_{i}=\frac{1}{2\pi}\oint_{\Gamma_{i}}\epsilon_{ab}\,n^{a}dn^{b},
\qquad
n^{a}=\frac{\phi^{a}}{\|\phi\|},
\label{eq:winding}
\end{equation}
with $\Gamma_{i}$ a small contour around the $i$-th zero, and the total
topological charge is $W=\sum_{i}w_{i}$.

At the representative value $\tau=0.04$ the defect curve is crossed twice, once
on its rising branch and once on the falling one. Table~\ref{tab:defects}
collects the positions and winding numbers. The inner defect carries $w=+1$ and
lies in the locally stable region, the outer defect carries $w=-1$ and lies in
the unstable region, and the total charge is
\begin{equation}
W=(+1)+(-1)=0
\label{eq:W}
\end{equation}
for every correction strength examined. The correction acts on one member of
the pair only. As $\eta$ runs from $0$ to $5$ the inner defect moves from
$\rh/|Q|=1.195657$ to $1.127639$, a shift of about $5.7\%$, while the outer
defect stays at $3.686025$ to six decimal places. The reason is visible in
Eq.~\eqref{eq:defect}: at the outer zero one has $\lambda x^{3}\simeq24.7$, so
$e^{-\alpha S_{0}}\sim10^{-11}$ and the correction is numerically absent. The
exponential term is a near-horizon effect and behaves like one.

\begin{figure}[ht!]
\centering
\includegraphics[width=\textwidth]{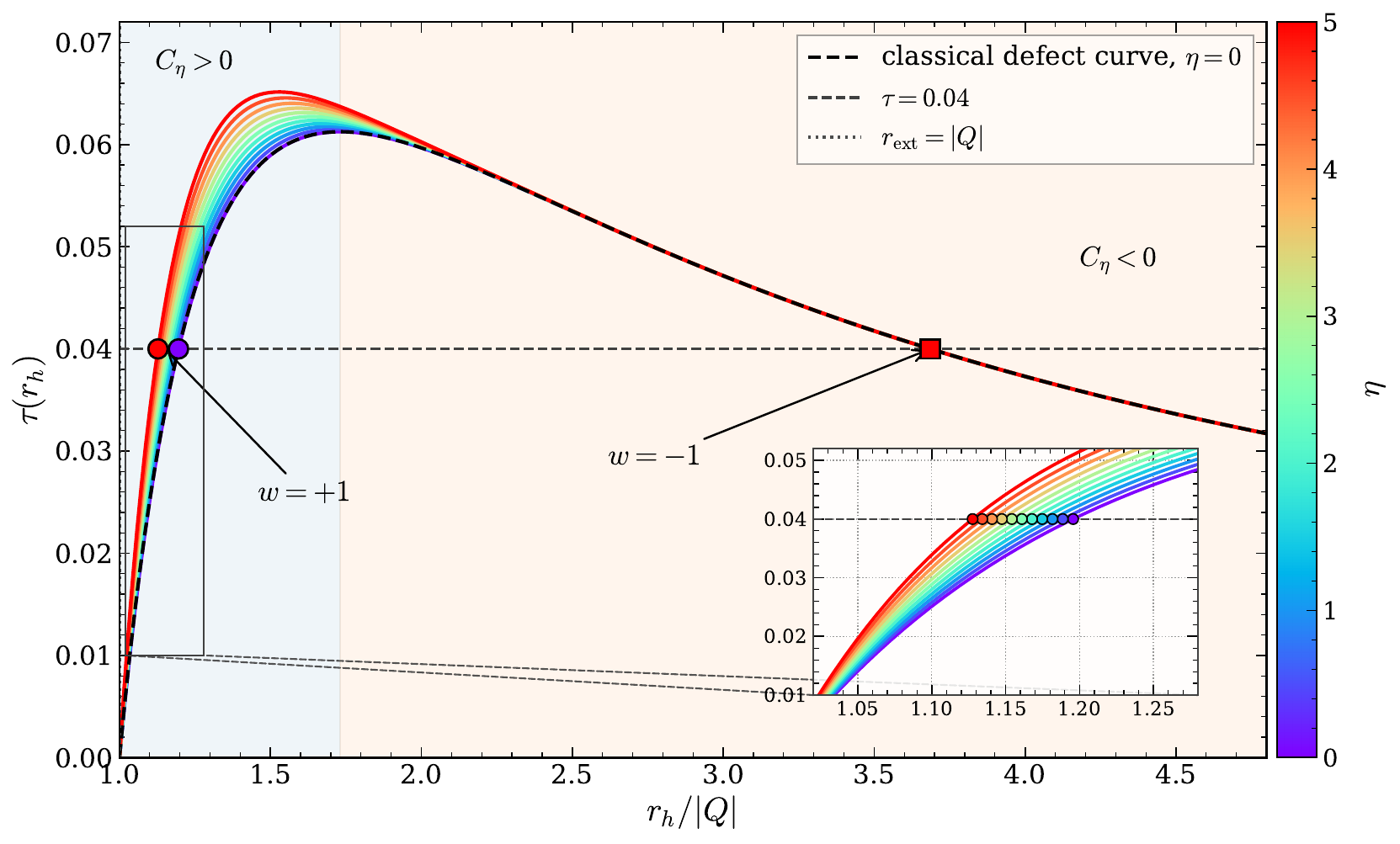}
\caption{Defect curves $\tau(\rh)$ of Eq.~\eqref{eq:defect} at $G_{5}=1$,
$Q=1$ and $\alpha=0.1$, with the correction strength carried by the colour bar
and the classical curve dashed. The horizontal line marks the auxiliary value
$\tau=0.04$ used in Table~\ref{tab:defects}; its two intersections with the
defect curve are the zeros of $\phi^{\rh}$, drawn as a circle on the rising
branch and a square on the falling branch. Background shading separates the
locally stable region from the unstable one at the classical transition radius.
The inset magnifies the boxed region around the inner zero, with guide lines
joining the box to the inset frame, and resolves the inward march of that zero
with increasing $\eta$; on this scale the outer zero does not move at all.}
\label{fig:defect}
\end{figure}

\begin{figure}[ht!]
\centering
\includegraphics[width=\textwidth]{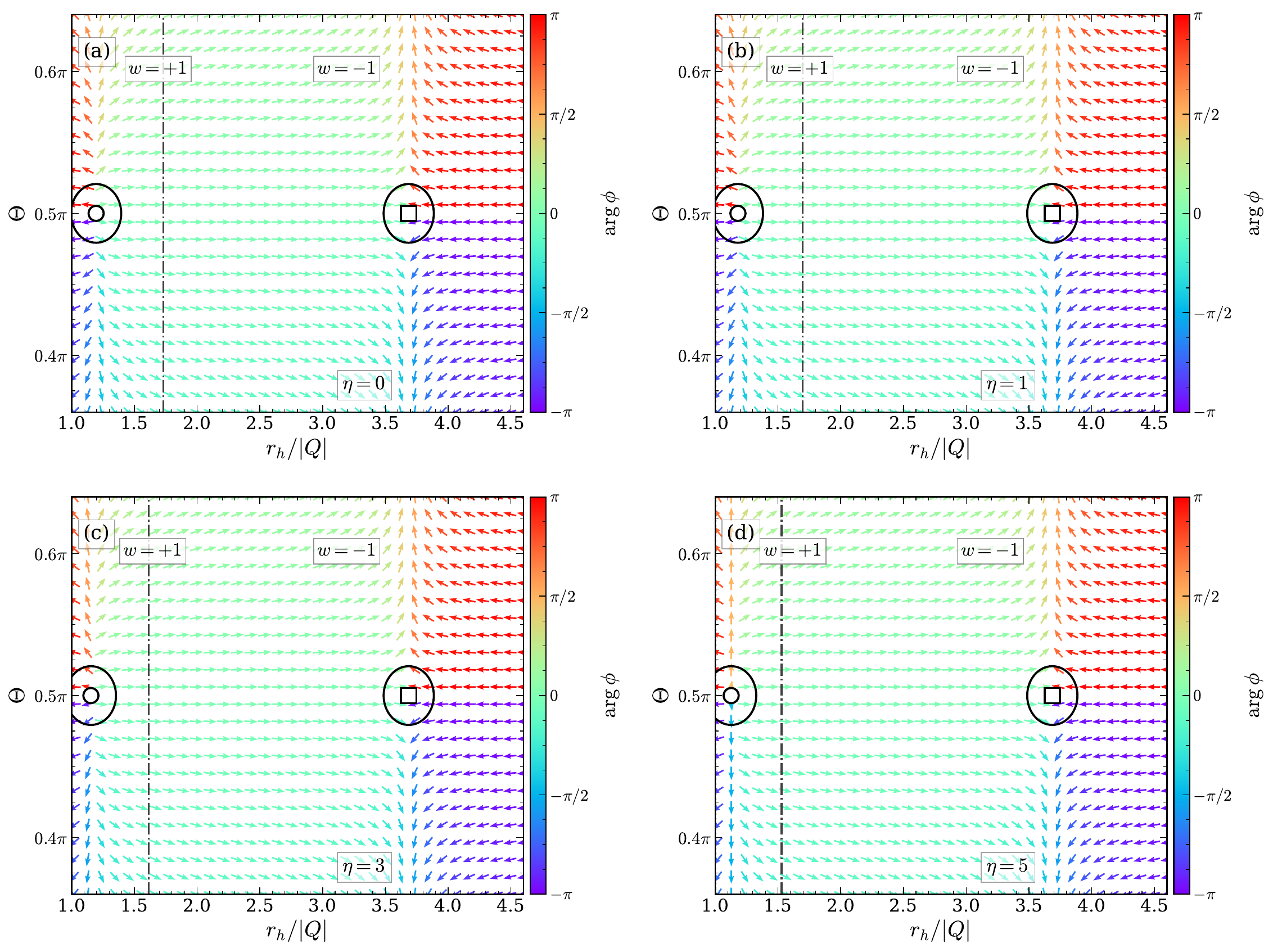}
\caption{Normalized vector field $\phi=(\partial\mathcal{F}_{\eta}/\partial
\rh,-\cot\Theta\csc\Theta)$ at $\tau=0.04$, $G_{5}=1$, $Q=1$ and $\alpha=0.1$,
for (a) $\eta=0$, (b) $\eta=1$, (c) $\eta=3$ and (d) $\eta=5$. Arrows are coloured by the local phase $\arg\phi$, so a defect of
winding $w$ is one around which the hue completes $|w|$ full cycles, in the
same sense as the colour bar for $w>0$ and the opposite sense for $w<0$. Each
zero is ringed by a contour of the type entering Eq.~\eqref{eq:winding}. The
dotted vertical line is the extremal radius and the dash-dotted line the
$\eta$-dependent Davies radius. Across the four panels the inner defect shifts inward while the outer one
holds its position, so the total charge $W=0$ is preserved throughout.}
\label{fig:flow}
\end{figure}

Figure~\ref{fig:defect} shows the defect curves and the displacement of the
inner zero, and Fig.~\ref{fig:flow} shows the normalized vector field with the
arrows coloured by $\arg\phi$, so that the $\pm1$ winding appears directly as
the number of times the hue wraps around each defect.

\begin{table}[ht!]
\setlength{\tabcolsep}{6pt}
\centering
\caption{Thermodynamic defects and winding numbers at fixed auxiliary parameter
$\tau=0.04$, with $G_{5}=1$, $Q=1$ and $\alpha=0.1$. Positions solve
$\partial\mathcal{F}_{\eta}/\partial \rh=0$, equivalently $\tau=\tau(\rh)$ from
Eq.~\eqref{eq:defect}. Every case carries an inner defect of winding $+1$ and
an outer defect of winding $-1$, so the total topological charge $W$ vanishes.
The correction displaces the inner defect inward and leaves the outer one
unchanged. The last column repeats the corresponding Davies-type radius.}
\label{tab:defects}
\begin{tabularx}{\textwidth}{CCCCCCC}
\hline\hline
\textbf{$\eta$} & \textbf{$\rh^{(1)}/|Q|$} & \textbf{$w_{1}$} &
\textbf{$\rh^{(2)}/|Q|$} & \textbf{$w_{2}$} & \textbf{$W$} &
\textbf{$r_{D}/|Q|$}\\
\hline
0 & 1.195657 & $+1$ & 3.686025 & $-1$ & 0 & 1.732051\\
1 & 1.181656 & $+1$ & 3.686025 & $-1$ & 0 & 1.696134\\
2 & 1.167819 & $+1$ & 3.686025 & $-1$ & 0 & 1.656968\\
3 & 1.154184 & $+1$ & 3.686025 & $-1$ & 0 & 1.615229\\
4 & 1.140782 & $+1$ & 3.686025 & $-1$ & 0 & 1.572021\\
5 & 1.127639 & $+1$ & 3.686025 & $-1$ & 0 & 1.528617\\
\hline\hline
\end{tabularx}
\end{table}

\section{The magnetic Yang--Mills perturbation sector}\label{isec6}

The scalar quasinormal spectrum of this geometry is already known. Guo and Miao
solved the Klein--Gordon equation on the five-dimensional EYM background and
mapped the charge dependence of the frequencies \cite{GuoMiao2020}, and the
eikonal correspondence with the unstable circular null geodesic was established
for the higher-dimensional family shortly afterwards \cite{GuoMiao2020b}. We do
not repeat the scalar effective potential, the WKB expansion
\cite{Konoplya2003,MatyjasekOpala2017,KonoplyaZhidenkoZinhailo2019} or the
photon-sphere analysis.

A probe scalar, however, answers a different question from the one we want. It
propagates on a frozen background and is blind to whether the Yang--Mills
configuration that sources that background is itself stable. In four dimensions
the answer for colored black holes was negative
\cite{StraumannZhou1990,Bizon1991,BrodbeckStraumann1996}, so the question is
worth asking in five. We therefore perturb the gauge field directly.

\subsection{Master equation and effective potential}\label{isec6a}

Okuyama and Maeda formulated the spherically symmetric magnetic perturbation
sector for the five-dimensional Einstein--Yang--Mills system
\cite{OkuyamaMaeda2003}. At vanishing cosmological constant their analytic
magnetic solution has a vanishing static Yang--Mills function, $w_{0}=0$, and a
logarithmic mass function, which is precisely the geometry of
Eq.~\eqref{eq:fh}. Perturbing about $w_{0}=0$ with $w=w_{1}(r)e^{-i\omega t}$,
the linearized Einstein equations return the metric perturbations
algebraically in terms of $w_{1}$; substituting them back into the linearized
Yang--Mills equation removes the metric perturbations entirely. With the master
variable and tortoise coordinate
\begin{equation}
\chi(r)=r^{1/2}w_{1}(r),
\qquad
\frac{dr_{*}}{dr}=\frac{1}{f(r)},
\label{eq:master}
\end{equation}
the system reduces to a single Schr\"odinger-type equation
\begin{equation}
-\frac{d^{2}\chi}{dr_{*}^{2}}+\VYM(r)\,\chi=\omega^{2}\chi,
\label{eq:schrodinger}
\end{equation}
with the effective potential, for the asymptotically flat case and with the
Yang--Mills charge scale restored,
\begin{equation}
\VYM(r)=\frac{f(r)}{4r^{4}}
\left[5\rh^{2}-4Q^{2}+10Q^{2}\ln\!\left(\frac{r}{\rh}\right)-9r^{2}\right].
\label{eq:V}
\end{equation}
Equation~\eqref{eq:V} vanishes at the horizon, where $f(\rh)=0$. Far away it
approaches
\begin{equation}
\VYM(r)=-\frac{9}{4r^{2}}
+\mathcal{O}\!\left(\frac{\ln r}{r^{4}}\right),
\qquad r\to\infty,
\label{eq:Vasym}
\end{equation}
which is negative. The potential is therefore not positive definite, and no
positivity argument can establish stability, exactly as found in the original
five-dimensional analysis \cite{OkuyamaMaeda2003}. A direct eigenvalue
calculation is unavoidable, and as Sec.~\ref{isec6d} shows, the coefficient
$9/4$ in Eq.~\eqref{eq:Vasym} does considerably more than block a positivity
proof.

\subsection{Dimensionless form and boundary conditions}\label{isec6b}

\begin{figure}[ht!]
\centering
\includegraphics[width=\textwidth]{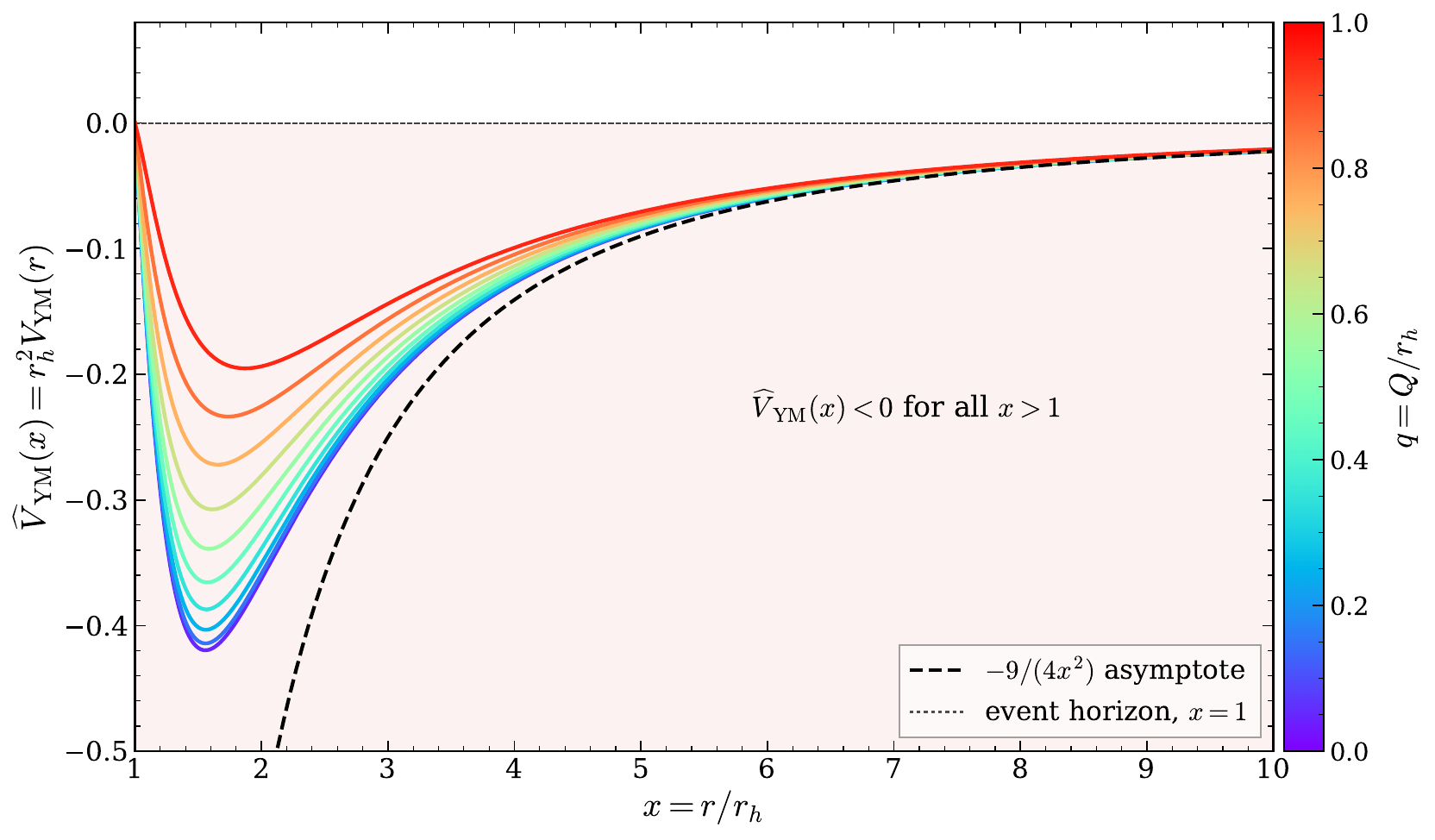}
\caption{Dimensionless effective potential $\Vh(x)=\rh^{2}\VYM(r)$ of
Eq.~\eqref{eq:Vhat} for the spherically symmetric magnetic Yang--Mills
perturbation, with the dimensionless charge $q=Q/\rh$ carried by the colour
bar. The event horizon is at $x=1$, where the potential vanishes. Raising the
charge makes the well shallower and moves its minimum outward, but the
potential stays negative throughout the exterior for every charge shown. The
dashed curve is the $-9/(4x^{2})$ tail of Eq.~\eqref{eq:Vasym}, which is
independent of $q$ and controls the accumulation of overtones discussed in
Sec.~\ref{isec6d}.}
\label{fig:potential}
\end{figure}

For the numerical work we set
\begin{equation}
x=\frac{r}{\rh},
\qquad
q=\frac{Q}{\rh},
\qquad
\widehat{\omega}=\omega \rh,
\qquad
\widehat{r}_{*}=\frac{r_{*}}{\rh},
\label{eq:dimless}
\end{equation}
so that the nonextremal branch is $|q|<1$ and the extremal configuration is
approached as $|q|\to1^{-}$, equivalently $\rh\to|Q|^{+}$. The metric function
and effective potential become
\begin{equation}
f(x)=1-\frac{1+2q^{2}\ln x}{x^{2}},
\qquad
\Vh(x)\equiv \rh^{2}\VYM(r)
=\frac{f(x)}{4x^{4}}\left[5-4q^{2}+10q^{2}\ln x-9x^{2}\right],
\label{eq:Vhat}
\end{equation}
and the master equation reads
\begin{equation}
-\frac{d^{2}\chi}{d\widehat{r}_{*}^{2}}+\Vh(x)\,\chi
=\widehat{\omega}^{2}\chi.
\label{eq:masterdimless}
\end{equation}
Figure~\ref{fig:potential} plots $\Vh$ across the nonextremal range. The
potential is negative everywhere outside the horizon for every charge examined.
Increasing $q$ makes the well shallower and pushes its minimum outward, but
never lifts the potential above zero, and the asymptotic tail
\eqref{eq:Vasym} is independent of $q$ altogether.

With the convention $e^{-i\omega t}$ the quasinormal boundary conditions are
$\chi\sim e^{-i\omega r_{*}}$ at the horizon and $\chi\sim e^{+i\omega r_{*}}$
at infinity. The negativity of $\Vh$ singles out a simple subclass, the purely
imaginary modes. Writing
\begin{equation}
\omega=i\Gamma,\qquad \Gamma>0,
\label{eq:imaginary}
\end{equation}
Eq.~\eqref{eq:masterdimless} turns into the bound-state problem
\begin{equation}
\left[-\frac{d^{2}}{d\widehat{r}_{*}^{2}}+\Vh(x)\right]\chi=-\Gamma^{2}\chi,
\label{eq:boundstate}
\end{equation}
whose boundary conditions are exponential decay at both ends,
$\chi\sim e^{\Gamma\widehat{r}_{*}}$ near the horizon and
$\chi\sim e^{-\Gamma\widehat{r}_{*}}$ at large radius. Negative eigenvalues of
the Schr\"odinger operator in Eq.~\eqref{eq:boundstate} are therefore in
one-to-one correspondence with unstable modes, and a positive imaginary part of
$\omega$ means exponential growth in time.

\subsection{Numerical spectrum}\label{isec6c}

We solve Eq.~\eqref{eq:boundstate} by two independent methods and use their
agreement as the error estimate.

The first is direct diagonalization. We integrate the background from
$dx/d\widehat{r}_{*}=f(x)$ on a uniform tortoise grid, discretize the
one-dimensional Schr\"odinger operator to second order in the grid spacing,
impose Dirichlet conditions on the numerically decayed master function at both
ends, and diagonalize the resulting symmetric tridiagonal matrix. Negative
eigenvalues $E_{n}<0$ convert to growth rates through
$\Gamma_{n}=\sqrt{-E_{n}}$.

The second is a Riccati shooting scheme, which avoids the exponential range
problem that afflicts naive two-point shooting over a long tortoise interval.
Setting $u=f\,\chi'/\chi$, Eq.~\eqref{eq:boundstate} becomes
\begin{equation}
f(x)\,\frac{du}{dx}=\left[\Vh(x)+\Gamma^{2}\right]-u^{2}.
\label{eq:riccati}
\end{equation}
The horizon is a fixed point of Eq.~\eqref{eq:riccati} at $u=\Gamma$, so the
inner boundary condition must be started from its first-order expansion. With
$x=1+\epsilon$ and $f'(1)=2(1-q^{2})$ one finds
\begin{equation}
u(1+\epsilon)=\Gamma
-\frac{f'(1)\left(1+q^{2}\right)}{f'(1)+2\Gamma}\,\epsilon
+\mathcal{O}(\epsilon^{2}),
\label{eq:riccatiBC}
\end{equation}
while at large $x$ the decaying solution gives
$u\to-\Gamma+9/(8\Gamma x^{2})$. Integrating inward and outward and matching
$u$ at $x=2$ fixes $\Gamma$. Because the fundamental mode has no node, $u$
stays finite throughout and the scheme is well conditioned.

Table~\ref{tab:qnm} reports the fundamental growth rate from both methods
across the nonextremal range. The two schemes agree to
$1.2\times10^{-7}$ or better at every charge, and the residual is dominated by
the second-order truncation of the finite-difference operator, as confirmed by
Richardson extrapolation in the grid spacing. The lowest eigenvalue is also
insensitive to the computational domain: shifting the outer boundary from
$\widehat{r}_{*}=40$ to $\widehat{r}_{*}=2000$ leaves $\Gamma_{0}$ unchanged in
the sixth decimal place.

The spectrum is unstable everywhere on the nonextremal branch. At the
thermodynamically distinguished radius $\rh=\sqrt{3}|Q|$, that is $q=1/\sqrt3$,
\begin{equation}
\widehat{\omega}_{0}=\omega_{0}\rh\simeq0.392322\,i,
\label{eq:daviesmode}
\end{equation}
and the growth rate falls monotonically with charge, from
$\Gamma_{0}\rh=0.414753$ in the $q\to0$ limit, where the potential reduces to
its Schwarzschild--Tangherlini form, to $0.317520$ at $q=0.99$. The decrease is
gentle: over the whole nonextremal range the rate drops by less than a quarter.
This weakening toward extremality should not be read as stability at
extremality. It says only that the dimensionless growth rate diminishes as
$|Q|/\rh\to1^{-}$, and the exact extremal point lies outside the reach of the
present scheme because the surface gravity vanishes there and the tortoise
coordinate degenerates.

The comparison with the thermodynamics is the point of Fig.~\ref{fig:gamma}(a).
The Davies scale $q=1/\sqrt3$ is where $C_{Q}$ diverges and the local stability
of the horizon reverses. The instability rate notices nothing. It varies
smoothly through that charge, with no kink, no divergence and no change of
slope beyond the smooth trend already present on either side. Within this
perturbation sector the Davies-type transition carries no dynamical signature.

\begin{table}[ht!]
\centering
\caption{Fundamental unstable mode of the spherically symmetric magnetic
Yang--Mills sector, written as
$\widehat{\omega}_{0}=\omega_{0}\rh=i\,\Gamma_{0}\rh$ with $\Gamma_{0}>0$. The
second column is the finite-difference result on a grid of $16001$ points over
$\widehat{r}_{*}\in[-45,200]$, the third the independent Riccati shooting
result, and the fourth their difference. The row $q=1/\sqrt3$ is the
Davies-type scale $\rh=\sqrt{3}|Q|$, and the first row is the
Schwarzschild--Tangherlini limit of the potential.}
\label{tab:qnm}
\begin{tabularx}{\textwidth}{CCCC}
\hline\hline
\textbf{$q=Q/\rh$} & \textbf{$\Gamma_{0}\rh$ (finite difference)} &
\textbf{$\Gamma_{0}\rh$ (shooting)} & \textbf{difference}\\
\hline
$\to 0$              & 0.41475258 & 0.41475247 & $1.14\times10^{-7}$\\
$0.05$               & 0.41460925 & 0.41460914 & $1.14\times10^{-7}$\\
$0.10$               & 0.41417731 & 0.41417720 & $1.13\times10^{-7}$\\
$0.20$               & 0.41242008 & 0.41241997 & $1.08\times10^{-7}$\\
$0.30$               & 0.40938223 & 0.40938213 & $1.01\times10^{-7}$\\
$0.40$               & 0.40488359 & 0.40488350 & $9.15\times10^{-8}$\\
$0.50$               & 0.39863427 & 0.39863419 & $7.96\times10^{-8}$\\
$1/\sqrt3\simeq0.5774$ & 0.39232194 & 0.39232187 & $6.94\times10^{-8}$\\
$0.65$               & 0.38492835 & 0.38492830 & $5.92\times10^{-8}$\\
$0.70$               & 0.37883389 & 0.37883384 & $5.21\times10^{-8}$\\
$0.80$               & 0.36349146 & 0.36349142 & $3.83\times10^{-8}$\\
$0.90$               & 0.34252255 & 0.34252252 & $2.66\times10^{-8}$\\
$0.95$               & 0.32936329 & 0.32936327 & $2.22\times10^{-8}$\\
$0.99$               & 0.31751959 & 0.31751957 & $1.94\times10^{-8}$\\
\hline\hline
\end{tabularx}
\end{table}

\subsection{An accumulating tower of unstable modes}\label{isec6d}

The fundamental mode is not the whole spectrum. The asymptotic form
\eqref{eq:Vasym} is an attractive inverse-square tail,
$\Vh\simeq-c/\widehat{r}_{*}^{2}$ with $c=9/4$, and the bound-state problem for
such a tail is governed by the sign of $c-1/4$. Writing the tail as
$\ell(\ell+1)/\widehat{r}_{*}^{2}$ with $\ell(\ell+1)=-c$ gives
$\ell=-1/2\pm is$ and
\begin{equation}
s=\sqrt{c-\tfrac14}=\sqrt{\tfrac94-\tfrac14}=\sqrt2,
\label{eq:s}
\end{equation}
which is real. This is the classic supercritical case
\cite{Case1950,FrankLandSpector1971}: the effective radial exponent becomes
imaginary, the operator is unbounded below in the absence of a short-distance
cutoff, and the discrete spectrum accumulates geometrically at zero energy. The
horizon supplies the cutoff here, so the fundamental mode is finite and well
defined, but the tower above it is not truncated. The eigenvalues satisfy
$E_{n}\propto e^{-2\pi n/s}$ at large $n$, hence
\begin{equation}
\frac{\Gamma_{n+1}}{\Gamma_{n}}\;\longrightarrow\;e^{-\pi/s}
=e^{-\pi/\sqrt2}\simeq0.108453 .
\label{eq:ratio}
\end{equation}
The ratio is fixed by the coefficient of the tail alone. Since that coefficient
is $9/4$ for every $q$, the accumulation rate is independent of the Yang--Mills
charge even though the individual $\Gamma_{n}$ are not.

\begin{figure}[ht!]
\centering
\includegraphics[width=\textwidth]{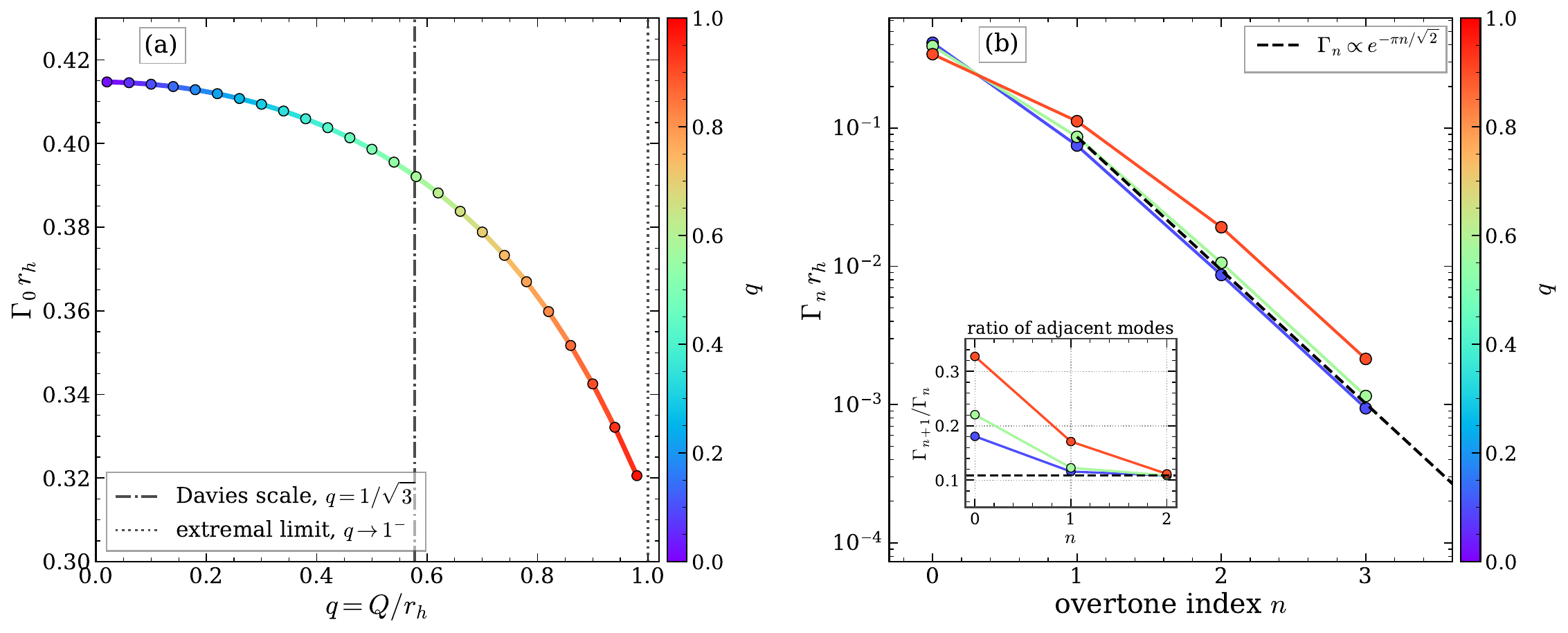}
\caption{Unstable spectrum of the magnetic Yang--Mills sector. Panel (a) is the
dimensionless growth rate $\Gamma_{0}\rh$ of the fundamental mode against the
charge ratio $q=Q/\rh$, with $q$ carried by the colour bar; markers are the
validated values of Table~\ref{tab:qnm}. The dash-dotted line is the
Davies-type scale $q=1/\sqrt3$ and the dotted line the extremal limit
$q\to1^{-}$. The rate falls monotonically toward extremality and passes through
the thermodynamic transition without any feature. Panel (b) shows the first
four members of the unstable tower at three charges on a logarithmic scale,
with the dashed line giving the asymptotic law
$\Gamma_{n}\propto e^{-\pi n/\sqrt2}$ of Eq.~\eqref{eq:ratio} anchored in the
regime where it applies. The inset is not a magnified window on the main
panel but a derived quantity: it plots the successive ratios
$\Gamma_{n+1}/\Gamma_{n}$ against overtone index on a guide grid. All three
charges converge to the same horizontal dashed asymptote $e^{-\pi/\sqrt2}$, as
expected for a constant driven by the tail of the potential alone.}
\label{fig:gamma}
\end{figure}

Table~\ref{tab:tower} tests this. At three representative charges the first
four growth rates were computed on domains extended to
$\widehat{r}_{*}=8000$, which is required because each successive overtone is
spatially an order of magnitude broader than the last. The successive ratios
approach Eq.~\eqref{eq:ratio} from above, reaching $0.1086$, $0.1088$ and
$0.1116$ by $n=2$ at $q=0.10$, $1/\sqrt3$ and $0.90$ respectively, in agreement
with the predicted $0.108453$ to the accuracy the domain permits. Pushing the
outer boundary to $\widehat{r}_{*}=3\times10^{4}$ resolves a fifth mode at
$q=1/\sqrt3$ and gives the next ratio as $0.1075$. Figure~\ref{fig:gamma}(b)
displays the tower and the convergence of the ratios.

Two remarks are in order. First, only $\Gamma_{0}$ is a domain-independent
number; the higher members are physical but require progressively larger
domains to resolve, and any truncation of the exterior region artificially
terminates the tower. Second, the accumulation does not make the instability
worse in any practical sense. The growth is dominated by $\Gamma_{0}$, and the
overtones are slower by an order of magnitude at each step. Their significance
is structural: the magnetic sector of this geometry has no gap above the
fundamental unstable mode, which is a property of the long-range tail rather
than of the horizon.

\begin{table}[ht!]
\setlength{\tabcolsep}{6pt}
\centering
\caption{Growth rates $\Gamma_{n}\rh$ of the first four unstable modes at three
charges, computed on a tortoise domain extending to
$\widehat{r}_{*}=8000$, together with the successive ratios
$\Gamma_{n+1}/\Gamma_{n}$. The ratios approach the charge-independent
asymptotic value $e^{-\pi/\sqrt2}\simeq0.108453$ predicted by
Eq.~\eqref{eq:ratio} from the inverse-square tail of the effective potential.}
\label{tab:tower}
\begin{tabularx}{\textwidth}{CCCCCCCC}
\hline\hline
\textbf{$q$} & \textbf{$\Gamma_{0}\rh$} & \textbf{$\Gamma_{1}\rh$} &
\textbf{$\Gamma_{2}\rh$} & \textbf{$\Gamma_{3}\rh$} &
\textbf{$\Gamma_{1}/\Gamma_{0}$} & \textbf{$\Gamma_{2}/\Gamma_{1}$} &
\textbf{$\Gamma_{3}/\Gamma_{2}$}\\
\hline
$0.10$              & 0.4141778 & 0.0748221 & 0.0086664 & 0.0009412
                    & 0.1807 & 0.1158 & 0.1086\\
$1/\sqrt3$          & 0.3923223 & 0.0864727 & 0.0106115 & 0.0011546
                    & 0.2204 & 0.1227 & 0.1088\\
$0.90$              & 0.3425227 & 0.1122088 & 0.0191906 & 0.0021407
                    & 0.3276 & 0.1710 & 0.1116\\
\hline\hline
\end{tabularx}
\end{table}

\subsection{Scope of the calculation}\label{isec6e}

Two limitations should be stated plainly. The calculation perturbs the magnetic
Yang--Mills configuration and eliminates the accompanying metric perturbations
through the linearized field equations, so it is a genuine stability test of
the gauge sector rather than a probe-field spectrum. It is not, however, the
full quasinormal spectrum of the higher-dimensional Wu--Yang system. A complete
treatment would decompose tensor, vector, scalar and internal-gauge
perturbations of the $SO(4)$ configuration in the Kodama--Ishibashi formalism
\cite{KodamaIshibashi2003,KodamaIshibashi2004,KodamaIshibashi2011}, which is a
considerably heavier problem and is left for future work. Other sectors could
in principle carry different stability properties, though the four-dimensional
precedent \cite{StraumannZhou1990,Bizon1991,BrodbeckStraumann1996} does not
encourage optimism.

The second limitation concerns the entropy correction. Section~\ref{isec4}
treats $S_{\eta}$ as a modification of the entropy at fixed metric, so the
perturbation problem solved here is the classical one. A self-consistent
treatment would correct the field equations and hence the potential
\eqref{eq:V} itself \cite{aram2}, and only then could the effect of $\eta$ on
$\Gamma_{0}$ be quantified. Since the correction is exponentially confined to
the near-horizon region while $\Gamma_{0}$ is set by the well between the
horizon and $x\simeq2$, we expect the effect to be small but do not claim it.

\section{Discussion and conclusions}\label{isec7}

We have assembled a description of the five-dimensional Einstein--Yang--Mills
black hole of Mazharimousavi and Halilsoy \cite{MazharimousaviHalilsoy2008}
that carries exact thermodynamics, a non-perturbative entropy correction, a
global topological classification, and a direct test of the stability of the
non-Abelian sector.

At the classical level the analysis closes in elementary functions. The horizon
condition gives the mass--radius relation \eqref{eq:M}, the temperature
collapses to the compact form \eqref{eq:T}, the extremal radius is
$r_{\star}=|Q|$ and the extremal mass follows in closed form. The area law
supplies the entropy, the first law is satisfied identically, and the conjugate
Yang--Mills potential \eqref{eq:Phi} completes the set. Two corrections to the
existing literature emerged along the way. The Gibbs-like potential is
Eq.~\eqref{eq:G}, in which the logarithmic term changes sign relative to the
Helmholtz free energy, and the corrected heat capacity carries the prefactor
$3\pi(\rh^{2}-Q^{2})/(4G_{5}\rh)$ of Eq.~\eqref{eq:Ceta}. We also recorded the
curvature invariants, Eqs.~\eqref{eq:ricci} and \eqref{eq:kretschmann}, which
show that the logarithm in the metric function strengthens rather than
regulates the central singularity.

The classical heat capacity diverges at $\rh=\sqrt3|Q|$, a Davies-type point
\cite{Davies1989} whose interpretation has been reconsidered recently
\cite{Davies2024}. Adding the exponential correction \eqref{eq:Seta} moves that
point inward, and the weak-correction expansion \eqref{eq:shift} gives the
displacement in closed form. At $\alpha=0.1$ the shift reaches almost $12\%$ by
$\eta=5$. This behavior parallels what has been found in logarithmic nonlinear
electrodynamics backgrounds \cite{aram3}, dilaton gravity \cite{aram4},
AdS--Rindler geometries \cite{aram5} and Lifshitz black holes \cite{aram6}: the
exponential term reorganizes the small-horizon sector and leaves the
large-horizon thermodynamics intact.

The topological description makes that statement precise. The defect equation
\eqref{eq:defect} turns out to be the corrected temperature itself, which ties
the winding-number classification directly to the local analysis. At fixed
$\tau$ we find a defect pair with $w=+1$ and $w=-1$ and total charge $W=0$,
unchanged for every correction strength examined, in line with the general
expectation that the topological number depends on the asymptotic
thermodynamics at small and large horizon radius rather than on the details in
between \cite{WeiLiuMann2022,WeiLiuWang2022}. What the correction does is move
the stable member of the pair while leaving the unstable member fixed to six
decimal places, because the exponential is suppressed by eleven orders of
magnitude at the outer defect. Non-perturbative quantum corrections deform the
near-horizon thermodynamics and preserve the global classification, which
matches the pattern reported for other entropy-corrected systems
\cite{aram3,aram4,aram5,aram6,GashtiPourhassanSakalli2025,GashtiSakalliPourhassan2025}.

The dynamical results are the part that does not follow from the
thermodynamics. Perturbing the magnetic Yang--Mills configuration itself, in
the framework of Okuyama and Maeda \cite{OkuyamaMaeda2003}, we found a purely
imaginary unstable mode across the whole nonextremal range, with
$\Gamma_{0}\rh$ falling monotonically from $0.414753$ at vanishing charge to
$0.317520$ at $q=0.99$. Two independent numerical schemes agree to better than
$1.2\times10^{-7}$. The mode passes through the Davies scale without any
feature, so within this sector a divergence of the heat capacity is not
accompanied by a divergent or even an anomalous dynamical timescale. That is
worth contrasting with holographic settings where quasinormal frequencies do
track thermodynamic transitions in the dual theory
\cite{KonoplyaZhidenko2011,HendiNemati2019}: the correspondence between
thermodynamic singularities and dynamical response is evidently more delicate
for intrinsic gauge-field perturbations than for probe fields.

The tower of Sec.~\ref{isec6d} was not anticipated. Because the effective
potential decays as $-9/(4r_{*}^{2})$ and the coefficient $9/4$ exceeds the
critical value $1/4$ for an inverse-square tail, the sector supports infinitely
many unstable modes accumulating at $\Gamma\to0$ with the ratio
$e^{-\pi/\sqrt2}\simeq0.108453$. Our numerics reproduce that constant at three
different charges, as they must, since it depends on the tail coefficient and
nothing else. The magnetic sector therefore has no spectral gap above its
fundamental instability, and the absence of a gap is a property of the
asymptotic region rather than of the horizon.

These results place the geometry in context. Hendi and Nemati studied
five-dimensional Yang--Mills massive gravity and reported van der Waals
behavior together with scalar WKB frequencies \cite{HendiNemati2019}, but for
planar horizons in the extended phase space; our setting is spherical, with a
logarithmic metric function, exponential entropy corrections and a topological
classification. Guo and Miao showed that the logarithm removes the divergence
of the Hawking temperature while preserving a charge-dependent scalar spectrum
\cite{GuoMiao2020,GuoMiao2020b}; the magnetic sector studied here answers the
different question of whether the solution holds together, and the answer is
that it does not. Pu and Yang found that regular EYM frequencies approach the
Schwarzschild values at large charge \cite{PuYang2023}, whereas the purely
imaginary mode found here persists across the entire charge range. Analyzing
gauge-field perturbations directly, rather than relying on scalar probes,
changes the conclusion.

A number of extensions follow naturally. Restoring a cosmological constant would
open the EYM--AdS phase structure, with Hawking--Page and van der Waals
behavior and a holographic reading of the entropy correction
\cite{HendiNemati2019,AliEtAl2024}. Moving to Einstein--Yang--Mills--Gauss--Bonnet
theory \cite{MazharimousaviHalilsoy2007} would show how higher-curvature terms
interact with the non-Abelian charge and the exponential correction. A
side-by-side comparison of logarithmic and exponential corrections in the same
background would clarify which scheme is appropriate in which regime
\cite{ChatterjeeGhosh2020,PourhassanEtAl2024}. The complete $SO(4)$ perturbation
problem remains open, as does the behavior of $\Gamma_{0}$ at exact extremality,
which needs numerical techniques adapted to vanishing surface gravity. Finally,
a self-consistent treatment in which the exponential correction back-reacts on
the metric \cite{aram2} would let the entropy correction and the instability
rate be computed within a single scheme, which is the natural next step for
this system.

\section*{Acknowledgments}
\.{I}.~S. thanks EMU, T\"{U}B\.{I}TAK, ANKOS and SCOAP3 for their support. He
also acknowledges COST Actions CA22113, CA21106, CA23130, CA21136 and CA23115
for their contributions to networking. The authors thank the anonymous referees
for comments that improved the manuscript.

\section*{Data Availability Statement}

The data supporting the findings of this study are contained within the article.
The numerical scripts used to generate the tables and figures, covering the
symbolic verification of the thermodynamic relations, the defect and
transition-radius roots, and the finite-difference and Riccati eigenvalue
solvers, are available from the corresponding author on reasonable request.

\bibliographystyle{unsrtnat}
\bibliography{final.ref}

\end{document}